\documentclass[conference]{IEEEtran}
\usepackage[nospace]{cite}

\usepackage[pdftex]{graphicx}
\ifCLASSINFOpdf
   \usepackage[pdftex]{graphicx}
   \graphicspath{./img}
   \DeclareGraphicsExtensions{.pdf,.jpeg,.png}
\else
\fi
\usepackage{array}
\usepackage{amsmath}

\newcommand{\ra}[1]{\renewcommand{\arraystretch}{#1}}

\usepackage{booktabs}

\usepackage[font=footnotesize,caption=false]{subfig}

\usepackage[table]{xcolor}

\usepackage{booktabs}
\usepackage{ulem}

\begin{document}
%
\title{On Designing and Testing \\Distributed Virtual Environments}

\author{
\IEEEauthorblockN{Arthur Valadares, Eugenia Gabrielova, Cristina Videira Lopes}
\IEEEauthorblockA{Bren School of Information and Computer Sciences, \\
University of California, Irvine\\
Irvine, CA, USA\\
Email: \{avaladar, eugenig, lopes\}@ics.uci.edu
}}


%


\maketitle




\begin{abstract} 

Distributed Real-Time (DRT) systems are among the most complex
software systems to design, test, maintain and evolve. The existence
of components distributed over a network often conflicts with
real-time requirements, leading to design strategies that depend on
domain- and even application-specific knowledge. Distributed Virtual
Environment (DVE) systems are DRT systems that connect multiple users
instantly with each other and with a shared virtual space over a
network. DVE systems deviate from traditional DRT systems in the
importance of the quality of the end user experience.

We present an analysis of important, but challenging, issues in the
design, testing and evaluation of DVE systems through the lens of
experiments with a concrete DVE, OpenSimulator. We frame our
observations within six dimensions of well-known design concerns:
correctness, fault tolerance/prevention, scalability, time
sensitivity, consistency, and overhead of distribution. Furthermore,
we place our experimental work in a broader historical context,
showing that these challenges are intrinsic to DVEs and suggesting
lines of future research.


\end{abstract}




\section{Introduction}

In software systems, the first measure of a successful design is the
fulfillment of functional requirements. However, functional
correctness is not enough; non-functional properties, i.e. the
operational characteristics, are equally important for the success of
software systems. In some systems, such as in Distributed Real-Time
(DRT) applications, non-functional requirements are often a critical
part of the overall function of those systems and need to be taken
into consideration from the early stages of design; neglecting
non-functional requirements can possibly render the software
useless. For instance, many social applications and online games, such
as Google Hangouts or Second Life, are naturally distributed, must
perform in real-time and must be resilient to failures. If the
response time between components of these systems is above a certain
threshold, or if the components fail systematically, these systems
become unusable.

Designing for distributed components and real-time responsiveness is
challenging, as these two requirements often hinder each
other. Distributed systems partition applications into independent
processes that can be deployed on separate hardware, communicating
through a network. The inter-process communication over the network
introduces a significant delay for real-time sensitive
applications. Thus, it is usually necessary to define a fine balance
between the desired level of distribution and real-time responsiveness
when designing a DRT application. 

But designing DRT systems is not the only difficult aspect of these
systems. Evaluating the success of those designs is also a non-trivial
task. It is often impractical, and clearly unwise, to evaluate and
test a DRT application as it is deployed in production. It is
impractical because the operation may require hundreds to thousands of
machines and users, and it is unwise because the application may not
be functioning correctly or may not be operating at an acceptable
level. It is then necessary to develop experiments and metrics that
can be expected to perform similarly to the production deployment. Yet
assumptions and abstractions of test deployments, such as unlimited
bandwidth, no jitter, and no thread context-switching costs, can be
made carelessly, resulting in unachievable performance in
production. Furthermore, choosing and interpreting the metrics that
demonstrate correctness and performance of a design also requires
careful consideration.

In spite of these difficulties, DRT systems become necessary when a
combination of properties from distributed systems and real-time is
required. The variety of DRT systems these days is very wide; this
paper focuses on one type of DRT system that we have more experience
with: Distributed Virtual Environments (DVE). DVEs are DRTs that
connect multiple users instantly with each other and with a shared
virtual space over a network. These environments have a broad range of
uses; applications range from shared observation of simulation of real
world physics, to games, to creating interactive platforms where users
can share experiences, engage in communication, and even modify the
virtual environment as they see fit. Examples of DVE applications
include World of Warcraft, Second Life, Google Hangouts, shared online
editors, advanced instant messaging systems (e.g. Slack), among many
others.

This paper presents an analysis of our experience with designing,
testing and evaluating a DVE, OpenSimulator~\cite{OpenSimulator}. In
doing that work, we have come across several challenges that, although
not new, illustrate very well the kinds of challenges that are present
in the development of DVEs. As such, the contribution of this paper is
twofold: (1) it provides a couple of concrete design and profiling
scenarios that are representative of a large spectrum of situations in
the development of DVEs; and (2) it reflects on those experiences,
placing them in an historical perspective of DVE and DRT research over
the years, showing that these challenges are intrinsic and quite
interesting as research topics.

The remainder of this paper is organized as
follows. Section~\ref{sec:prior} presents the context for the case
studies and their analysis. Sections~\ref{sec:design} and
\ref{sec:testing} present the two experiment case studies and the
lessons learned in each. Section~\ref{sec:related} places those
observations in an historical perspective. Finally,
\ref{sec:conclusion} offers some departing thoughts.

\section{Context and Prior Work}
\label{sec:prior}

\subsection{DVEs and OpenSimulator}
\label{subsec:dves}

The design of DVEs tends to fall into two camps:
peer-to-peer~\cite{keller2002toward,frey2008solipsis,maamar2010} and client-server
architectures~\cite{dionisio2013virtual}. Although peer-to-peer DVEs are
very popular in research, most commercial DVEs are done in a
client-server architectural style: users connect to a single
server[-side], responsible for maintaining rules, generating
reactions, and broadcasting updates to all users. The reasons for the
industry preference fall beyond the scope of this paper, but security
and privacy are some of the major concerns. Small DVEs are able to
serve all those functions from a single server. However, as the number
of shared objects and users increases, single servers are
bottlenecks~\cite{Almroth2010,Valadares2012,Brandt2005,Blackman2009}. The
responsiveness of these environments is highly dependent on the number
of real-time events that need to be distributed, and those events are
highly dependent on users' actions; in other words, performance of the
system, as a whole, is highly application-specific.

OpenSimulator~\cite{OpenSimulator} is an open-source virtual
environment framework that uses the same protocol as Second Life;
it is a clean-room reimplementation of the server-side of
Second Life that is able to use the unmodified Second Life
client. Comparable DVE open-source distributed simulator
implementations are OpenWonderland \cite{Kaplan2011} and Meru
\cite{Horn2010}, but these are less popular than OpenSimulator. Over
the past 8 years, we have been contributing to the development of
OpenSimulator.\footnote{The third author is one of the main core
  developers of OpenSimulator, and the other two authors have
  contributed code to it.} Particularly important to this paper are our
contributions for alternative architectures for scalability, and in
performance assessments of various scenarios that seem to be
problematic in real world usage of OpenSimulator.

Many scalability approaches for virtual environments involve space
partitioning techniques. In earlier space partitioning
methods~\cite{Carlsson1993,Waters1996}, space is partitioned in
fixed-size large areas of space, sometimes referred as
\textit{regions} or \textit{worlds}. Due to constraints dictated by
the client-server protocol, OpenSimulator inherited that architecture
from Second Life itself. In OpenSimulator, like in Second Life, the
world is divided in blocks of 256 meters squared, and each
\textit{region} is simulated on a different simulator server. A novel
and more flexible approach is space partitioning through
\textit{microcells}\cite{Vleeschauwer}, which are indivisible small areas of
space that can be grouped to form custom shaped partitions that better
adapt to load. But even specialized space partitioning methods alone
were shown to be insufficient under certain conditions of load
\cite{Liu2010}. Many other load partition schemes can be designed.

The \textbf{Distributed Scene Graph (DSG)} is a client-server
architecture for decoupling \textit{Scene} and \textit{operations}
\cite{Lake2010,Liu2010,Liu:2010} in OpenSimulator. Scene is the data
that represents the state of the virtual world, where operations are
responsible for reading and writing to the Scene. An example of Scene
state is an object's position and velocity. An example of an operation
is dropping the object from a certain height, and have physics
operations update its state over time. In DSG, multiple simulators
share and synchronize the Scene while each simulator can be dedicated
to independent groups of operations. DSG uses an eventually-consistent timestamp-based synchronization protocol for resolving updates between simulators. The clocks are synchronized using the Network Protocol Time (NTP) service, and the update with the highest timestamp is applied on every simulator. For more details on the consistency model used in DSG, see Liu et al. \cite{Liu2012}.

In \textbf{DSG with Microcells (DSG-M)} \cite{Valadares2014} we
redesigned DSG to push scalability further, by allowing simultaneous
decoupling of operations and space through \textit{microcell}
partitions~\cite{Vleeschauwer}. DSG-M allows for simulators to be
partitioned in both dimensions, enabling better adaptation to
load. DSG-M was evaluated through a physics intensive experiment, and
partitioning of both functionality (e.g. physics, script) and space,
by dividing the region space in half. When compared to DSG, DSG-M
results showed a 15\% improvement in performance in the worst-case
scenario and nearly double for a perfectly partitioned space scenario
(i.e. no inter-partition communication).

The case studies in this paper pertain to our experiments with
the design and evaluation of DGS-M, and to the evaluation of
specific problematic situations in unmodified OpenSimulator.

\subsection{Six Dimensions of Concern} \label{sec:properties}

The evaluation of the design, and the systematic testing of any DVE
require the existence of well defined metrics for establishing
acceptable behavior. In a previous paper~\cite{Valadares2014DSRT}, we
formulated six concerns that capture important tradeoffs of DRT
systems: correctness, fault tolerance, parallelism, time sensitivity,
consistency, and overhead costs. As such, in our OpenSimulator work,
we have used metrics in all of these dimensions of concern. 

The research community has long identified these, or variations of
these, as major concerns for these systems
(e.g.~\cite{stancovic1988misconceptions,cristian1991understanding,schutz1994fundamental,muir2004seven}).
We will give a more in-depth historical perspective on these issues in
Section~\ref{sec:related}. In order to ground our case studies, we
give just a brief description of each of these dimensions of concern.

\begin{itemize}
\item {\bf Correctness.} In a traditional algorithmic perspective,
  when an algorithm is correct, execution will produce correct results
  repeatedly. External factors, such as the operating environment, are
  abstracted away. However, even when ignoring the inevitability of
  application defects, most DRT systems are not deterministic because
  the operating environment is a fundamental piece of these
  systems. Hardware and physical devices such as CPU, memory, and
  networking can influence computation results due to exhaustion of
  resources or heterogeneity of hardware. Additionally, interactive
  systems, such as DVEs, process user input continually, and don't
  necessarily produce a final output. Determining the behavior to be
  correct requires more malleability, room for imprecision and, many
  times, a fair amount of ingenuity.

\item{\bf Fault tolerance.} Fault tolerance is a system's ability to
  survive failures, and is a highly desired property of distributed
  systems. Distributing computation adds more hardware and networking,
  increasing the chance of a single component failing. As a
  distributed system grows, so does chance of failure. Through fault
  tolerant design and algorithms, a distributed system can be made
  robust against individual component failures, typically at the cost
  of overhead resource usage in coordination, replication, and
  redundancy.

\item{\bf Parallelism and Scalability.} Parallelism enables
  computation to be partitioned and executed in parallel. Partitioning
  computation may require coordination, which may be required only at
  the start and end, at a certain rate during the execution, or not at
  all. Parallelism comes in multiple forms in software development.
  Distributed systems have networked parallelism, where processes are
  executed in different hardware, connected through a network. The
  advantage of parallelism is increasing computing power by adding
  more networked hardware resources, improving software
  \textit{scalability}. This form of scalability is referred as
  \textit{horizontal scalability}. 

\item{\bf Time Sensitivity.} DRT applications have real-time requirements,
  meaning they have time sensitive I/O. Time sensitivity can be
  originated from interactivity from users or from computation of
  other software components, as in a pipe and filter architectural
  style.

\item{\bf Consistency.} The consistency property determines how each
  participating node of the distributed system maintains shared
  state. Shared state can be always consistent, eventually consistent,
  or allow for inconsistency. Enforcing consistent states for every
  node at every point in time would require strong consistency
  algorithms that may break real-time requirements. Many DRT
  applications use \textit{eventual consistency}. Nodes in the DRT
  system will have slightly different states during execution, but
  state will eventually converge to the same values.

\item{\bf Overhead Costs.} DRT applications pay an overhead cost for
  distributing computation. Often the price is in network messaging
  and in coordination. Messages passed through the network stack can
  produce latencies from tens to hundreds of milliseconds. High
  frequency of messages can make latencies worse, and incur
  significant CPU usage for packing and unpacking
  messages. Coordination requires computing the partitions,
  distributing them through the network, and joining the results. When
  joining results, the coordinator must wait for all processes to
  respond, meaning the system will move at the speed of the slowest
  process. Different DRT applications have more or less sensitivity to
  overhead costs, depending on the degree of network messaging and
  coordination required.
\end{itemize}

Many of these concerns are hard to measure, as they are often
application-specific and are also correlated with multiple hardware
measurements such as CPU, memory, and latency. Evaluations of DRT
systems must account for multiple external factors, such as hardware,
network, and operating system. Furthermore, many DRT applications
cannot, or should not, be tested in production; thus, controlled
experiments must be used to mimic real-world usage.

\section{Case Study 1: DSG-M}
\label{sec:design}

This section presents a study of the evaluation of DSG-M, an extension
of DSG, which, in turn, is an extension of OpenSimulator. In designing
DSG-M we wanted to know whether, in practice, it was ``better'' or
``worse'' than DSG.\footnote{We had previously presented some results
  of DSG~\cite{Lake2010} comparing it to unmodified
  OpenSimulator.} We describe the rationale behind the experiments
and metrics, and conclude the section with observations regarding the
challenges we encountered.

\subsection{Objective}

In terms of the six dimensions of concern presented in the previous
section, DSG-M was designed to be \textit{parallel/scalable} above all
else, much more than the basic OpenSimulator and the DSG extension;
\textit{correctness} and \textit{time sensitivity} were secondary, but
important, concerns. Thus, the objective was to measure precisely the
systems' performance along those dimensions. In other words, the
evaluation goal was to assess how much more \textit{scalable} than DSG
DSG-M was under acceptable \textit{correctness} and \textit{time
  sensitivity} intervals. The secondary objectives were to identify
processing bottlenecks of the distributed simulation and to estimate
the computing power required per simulated entity.

\subsection{Experiment}

\begin{figure}[h]
\center
\includegraphics[width=\linewidth]{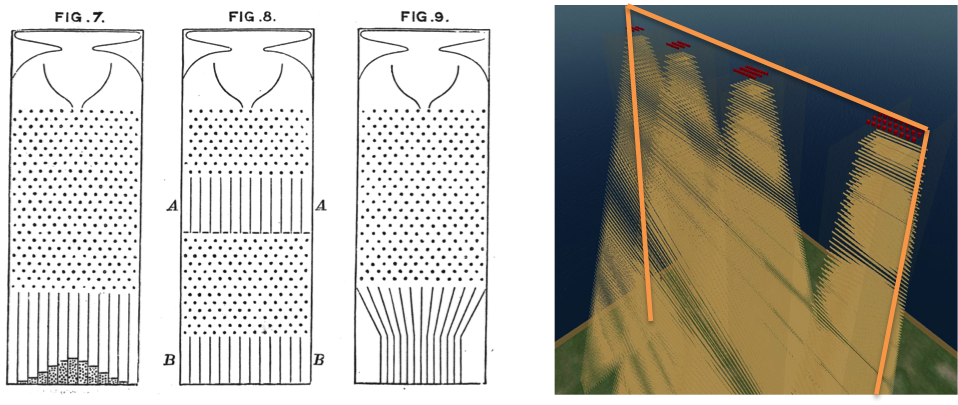}
\caption{On the left, the original drawing for a Galton box \protect\cite{galton1889natural}. On the right, the simulated Galton boxes for the experiment. The orange line shows how the virtual space was divided for experiments. The boxes at the top drop the balls.}
\label{fig:galton_box}
\end{figure}

After much consideration about how to measure behavior that could both
push the limits of the simulators and be ``correct,'' we designed a
physics-based experiment; we chose to simulate a device called
\textit{Galton box} \cite{galton1889natural}. Figure
\ref{fig:galton_box} shows both the original sketch of the device, and
the simulated Galton box. The Galton box is a board with multiple rows
of equally spaced pegs. Each row from top to bottom adds and extra peg
and is shifted, so that each peg is exactly in the middle of the gap
of pegs in the row above. At the bottom there are buckets covering the
gap between each two pegs. The device works by dropping balls at the
top of the box. Each time the ball falls on a peg, there is a 50\%
chance of it dropping to the right or to the left of the peg. If
multiple balls are thrown in the same fashion, the buckets in the
bottom will have a normal distribution of balls per bucket.

The Galton box experiment has many advantages for evaluating
DSG-M. First, it is easy to determine the expected behavior. At the
end of the simulation, the balls collected in the buckets should match
the binomial distribution. With a large enough number of balls, the distribution
becomes normal. Second, we can drop tens of thousands of balls, in order to obtain a
statistically significant and repeatable result. Third, to increase
load we simply drop the balls at a faster rate. Finally, the normal
distribution nature of the experiment allows us to test DSG-M under
worst-case conditions. Most of the balls will be crossing near the
middle of the device. If we partition the space so that the Galton box
is divided in half, we expected very high overhead costs in migrating
objects from a simulator to another.

\subsection{Setup}

The evaluation consisted of comparing the results of the experiments
on two system designs:

\textbf{A. Non-partitioned:} One simulator responsible for handling
the entire physics workload. This is the baseline case, corresponding
to DSG.

\textbf{B. Two Partitions:} Two simulators shared the workload of
computing physics by splitting the region in half. The split separates
each of the Galton Boxes in half, as shown by the orange lines in
Figure \ref{fig:galton_box}. As there are 3 rows of droppers being
split, one partition will contain one row, while the other will
contain 2. This setup had two sub-cases, active and passive
subscriptions, details of which are out of scope for this article. For
more information, refer to the original paper~\cite{Valadares2014}.

To generate enough load to overwhelm a single simulator process, we
used 4 Galton boxes of $n = 93$ levels, with 27 droppers each (3 rows
of 9). All droppers are at the exact same height and the 3 rows of
each Galton box are aligned across all Galton boxes. Droppers create
balls at an experiment-defined period of \textit{t} seconds per
ball. Each dropper drops 350 balls per experiment, at a configurable
rate \textit{t}. Each row is offset by a space larger than the
diameter of the ball, so there are nearly no collisions between
balls. By decreasing \textit{t}, balls are generated faster and
simulation load is increased.

\begin{figure}[h]
\center
\includegraphics[width=0.9\linewidth]{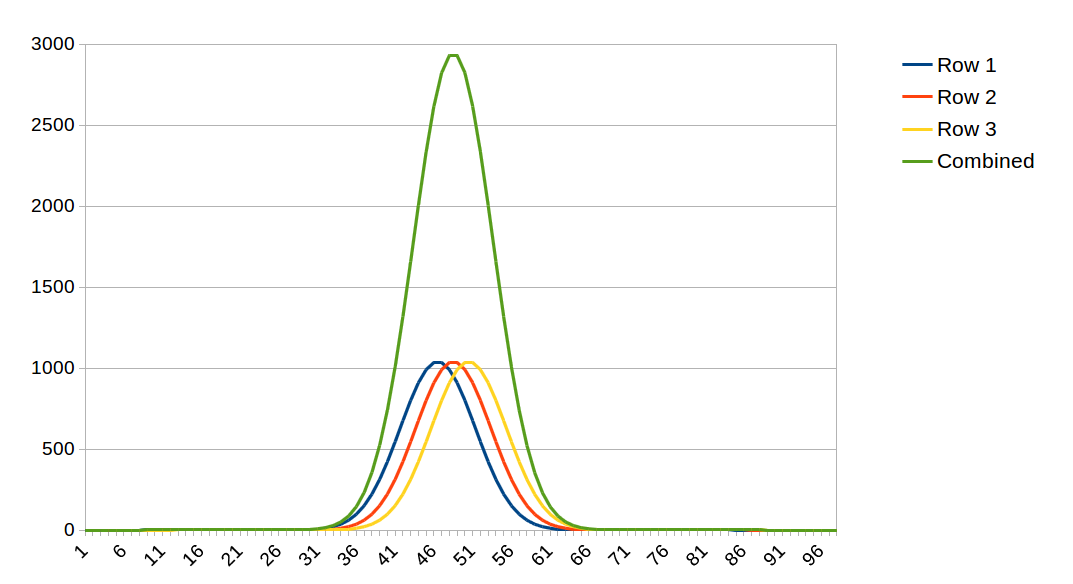}
\caption{Example distribution of 3 rows of droppers and the combined distribution for 37800 dropped balls. Notice that there are 93 levels, but 96 buckets, due to the overlapping of binomial distributions.}
\label{fig:theoretical_distribution}
\end{figure}

Considering the modified Galton box has 3 rows of droppers, each row
will generate a binomial distribution with a different average and
same standard deviation. Figure \ref{fig:theoretical_distribution}
shows the expected theoretical distribution of an experiment with
37,800 dropped balls. The Figure shows the 3 expected distributions
and their overlapping total.

Each simulator runs by itself on a dedicated desktop, connected by a Local Area Network. The desktops are Intel Core i7-2600 CPU @ 3.40 GHz, 16GB RAM, and 1Gbps Ethernet connections. The operating system is Ubuntu 12.10, and the simulators run on mono 3.2.8. 

\subsection{Metrics and Results}

The concerns we were interested in evaluating were
\textit{correctness}, represented as ball distribution in buckets,
\textit{time sensitivity}, by verifying whether or not the simulation
runs in real-time, and \textit{scalability}, represented as the
balance of load and performance, with and without space partitioning.

The metric for correctness is ball distribution per bucket. To compare
the results with the baseline, we used root mean square
error (RMSE).

For time sensitivity, we first opted for CPU as a measure of
performance. If the simulator is overwhelmed (high CPU), the
simulation will slow down, and real-time behavior will be lost. Later
on, we changed this metric to ball drop interval: the time a ball takes from creation at the top to destruction on the floor. This value is of $124.82 \pm 1.42$ seconds on normal conditions for DSG (i.e. non-overwhelmed simulator). The reason for
this change will be discussed in the next subsection. Other metrics
collected were number of messages exchanged for each simulator, number
of messages in the queue to be sent, and network bandwidth.

The scalability metric is simply the number of balls being
simulated. The more we can simulate, the better we can scale. The
experiments consists of 37,800 balls being dropped on the 4 Galton
boxes. Balls are created at a fixed period of \textit{t} balls per
second, and the experiment is repeated for different values of
\textit{t}. Any balls that do not fall within the boundaries of the
bins are discarded from the results.
The expectation was that dividing the region space by half would
enable simulation with a faster drop rate (i.e. higher load), and that
CPU\% would be perfectly correlated with the simulator increase in
load.

Figure \ref{fig:ex3} shows a summary of the results for physics
simulators for two experiments with the same period of ball generation
$t = 6$ seconds (i.e. 1 ball generated for every 6 seconds, for each
dropper). One experiment was partitioned by operation (i.e. one
physics simulator), but not by space. The second experiment was
partitioned by both operation and space, with two physics simulators
dividing the region in half. 

\subsection{Observations}

What follows is a list of the most relevant reflections from these
experiments that are relevant for DVE research.

\begin{figure*}%
\centering
\subfloat[][]{%
\label{fig:ex3-a}%
\includegraphics[width=0.4\linewidth]{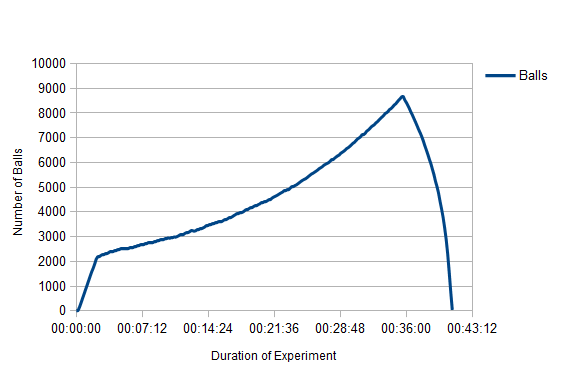}}%
\hspace{1pt}%
\subfloat[][]{%
\label{fig:ex3-b}%
\includegraphics[width=0.4\linewidth]{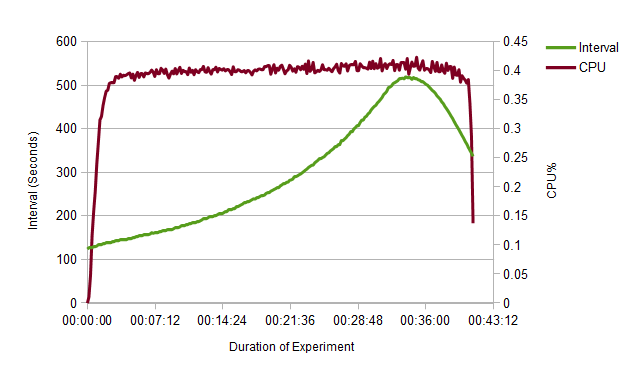}} \\
\subfloat[][]{%
\label{fig:ex3-c}%
\includegraphics[width=0.4\linewidth]{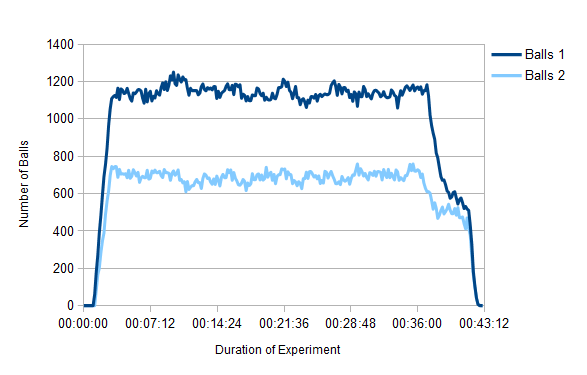}}%
\hspace{1pt}%
\subfloat[][]{%
\label{fig:ex3-d}%
\includegraphics[width=0.4\linewidth]{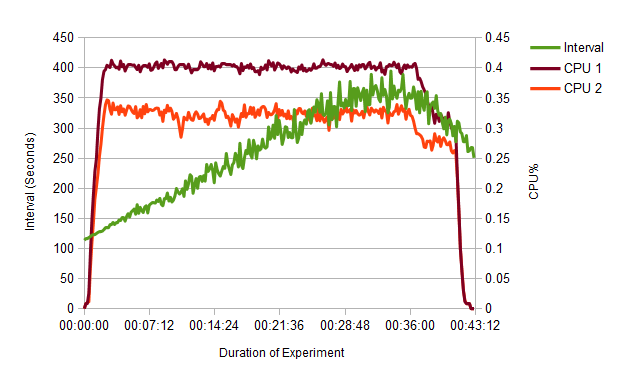}}\\
\caption[]{Number of balls (left), average interval between creation and collection (right), and CPU usage (right) over wall-clock time in 24 hour format. \\
\textbf{\subref{fig:ex3-a} and \subref{fig:ex3-b}}: Experiment A: one physics simulator;
\textbf{\subref{fig:ex3-c} and \subref{fig:ex3-d}}: Experiment B: two
physics simulators (1 and 2), dividing region in half, but with 2/3 of
the balls in one simulator and 1/3 in the other.}%
\label{fig:ex3}%
\end{figure*}

\noindent\textbf{1. Correctness.}

The experiments were designed to assess the differences in
\textit{scalability} between DSG and DSG-M under acceptable intervals
of \textit{correctness} and \textit{time sensitivity}. We knew what
kinds of design changes we wanted to try in DSG-M regarding
scalability, and we also knew how to measure time sensitivity; but we
needed some concept of correctness, and that was surprisingly not
trivial. Correctness in a DVE can be seen under two perspectives,
namely: (1) From an algorithmic perspective, functional correctness
is the primary goal; if the DVE has physics, for example, objects are
expected to drop with an acceleration similar to gravity; collisions
are expected to conserve momentum; if an object moves through a wall,
it is expected to be halted upon collision, etc.; and (2) From a user
perspective, all that matters is how believable or immersive the
virtual environment is;  incorrect behavior that cannot be noticed
by people is tolerated.

In designing the experiments, we faced the question of whether to
assess \textit{correctness} from the user perspective or from the
system perspective. A user-facing experiment would, in many ways, be
more meaningful, but doing user studies is a time-consuming effort
that requires either a large number of independent subjects or a large
time commitment on the part of a few. Plus, it is quite hard to design
meaningful perception metrics without a deep understanding of the
human visual system. Algorithmic experiments are much easier to
implement and measure. Physical simulations, in particular, have
properties that make them ideal for a precise evaluation. First,
physics results can be compared to real-world results for
correctness. Second, by not requiring users, tests can be performed
thousands of times, guaranteeing statistically significant results. We
ended up doing the Galton Box simulation, a form of assessing
correctness of a physics simulation adopted from experiments by
close collaborators~\cite{adams2011}.

However, it is important to be aware that algorithmic correctness is
not the same as user-level correctness, and this is an important
distinction that designers of DVEs need to take into consideration.

\noindent\textbf{2. Choice of baseline.}

As mentioned before, the metric for \textit{correctness} for this
experiment was the distribution of balls in bins. We derived
the theoretical predictions for that number of balls, and started by
using this prediction as the baseline; the initial approach was to
compare \textit{correctness} between DSG-M and DSG under heavy
workloads by comparing the empirical results of both with the expected
theoretical values. Trial runs showed a distribution that resembled a
normal curve, but further analysis showed that the standard
deviation of the distribution in the experiment in a non-stressing
scenario was higher than in theory. In an effort to understand the
deviation from theory, we realized the issue did not lie with DSG or
DSG-M, but rather with the physics simulation of OpenSimulator itself:
the physics engine was not precise enough to match the theoretical
expected distribution of balls to bins. In other words, we assumed
that the physical simulation under normal operating conditions matched
the theory, but that was not the case.

The solution was to replace the theoretical baseline with an empirical
one. Specifically, the new baseline came from measuring the 
results of DSG under a non-stressing workload. 

Relying on a theoretical prediction as baseline to compare two
architectures may seem like a perfectly reasonable approach, but it is
naive. In general, in DVEs it is hard to abstract away the influence
of the operating environment, which often distorts theoretical
predictions. Comparisons of alternative designs must always be
done with empirical baselines.

\noindent\textbf{3. Interpretation of metrics.}

Maintaining real-time behavior (i.e. \textit{time sensitivity})
through the experiments is essential in order to validate
\textit{scalability} results. However, operating system metrics such
as CPU, memory, and bandwidth, may not represent \textit{time
  sensitivity} appropriately, and, if used, may lead to
misinterpretations. In order to interpret them correctly, it is
essential to understand the internals of the software being assessed.

For example, it is tempting to assume CPU load is a measure of
processing load and that, eventually, an overwhelmed program will
reach 100\% CPU. However, modern CPUs are multicore, and many
applications these days are multi-threaded. In OpenSimulator, there
are two long-lived main threads (one of them being physics
simulation), and one additional thread per connected user. Some of
these threads may spawn several short-lived threads as they process
events from various sources; physics, however, does not spawn
threads. Hence, even when the physics simulation is overwhelmed, not
all cores are used; as a result, the maximum CPU\% on an 8-core
machine (as in our case) was never reached. In figures \ref{fig:ex3-b}
and \ref{fig:ex3-d}, CPU usage tops at around 40\%, independent of how
much the physics workload was increased. This result only makes sense
when there is a deep familiarity with the code of OpenSimulator,
specifically its concurrency model, which we briefly described here.

In the case of OpenSimulator and the extensions studied, the CPU
numbers reflected a combination of computing tasks, some of which were
from physics, some of which were not. As such, CPU usage did not quite
capture the performance issue we were looking to measure. In order to
isolate and measure the processing time allocated to physics, we
needed another metric; we used the time the balls take from creation
at the top to destruction on the floor. When physics is overwhelmed,
this interval increases.

In general, operating system metrics may not capture what is important
to measure. When those generic metrics are used, they need to be
interpreted according to in-depth knowledge of the software, or they
may be misleading. In many cases, application-specific metrics become
necessary.

\noindent\textbf{4. Dependent variables and masking.}

In both DSG and DSG-M, ball creation was done in the script simulator,
and ball deletion was done in the physics simulator(s). By design, the
script simulator was never overloaded on any experiment, and dropped
balls at a constant speed. The physics simulator(s) eventually became
overwhelmed with the number of balls, and slowed down all physics
operations, including object deletion. A slower simulator affected
\textit{time sensitivity}: the simulation started running slower than
real-time. This can be seen in Figures~\ref{fig:ex3-a} and
\ref{fig:ex3-b}; these Figures show an overwhelmed physics simulator in DSG
taking increasingly longer time to delete the balls (\ref{fig:ex3-b}),
which results in an increasing number of balls staying in the scene to
be physically simulated (\ref{fig:ex3-a}) -- a loop that produces
super-linear growth of both interval and number of balls, until the
simulator crashes.

In a non-distributed, single-threaded architecture this would not
have happened. Instead, a slow down of physics (and deletion of balls)
would also slow down the scripts (and creation of balls), resulting in
a constant number of balls to be physically simulated, even when the
simulator would be overloaded.

Interestingly, we observed a constant number of balls in the scenes
for the DSG-M experiment (Figure~\ref{fig:ex3-c}), where there were
two physics simulators, both of them overloaded
(Figure~\ref{fig:ex3-d}). At first, we interpreted this as a strongly
positive result for DSG-M: it looked like DSG-M was capable of
handling the load under stress much better than DSG. But this result
was puzzling: how could the experiment for two overloaded simulators
show such a different result from that with just one overloaded
simulator, considering that in both cases the data clearly showed an
increase in the interval between creation and deletion?  Where were
those ``zombie'' balls being processed?

\begin{figure}
\center 
\includegraphics[width=0.9\linewidth]{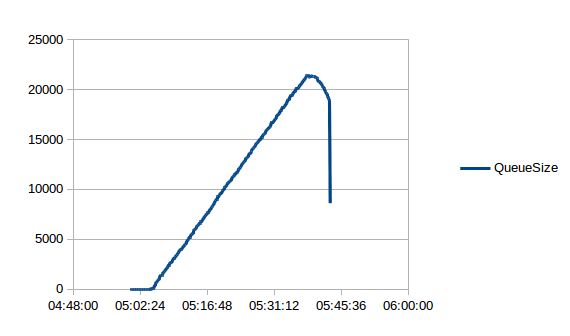} 
\caption{Message queue size intended for physics simulator over time in partitioned experiment.} \label{fig:queuesize} 
\end{figure}

The culprit behind this puzzling result was the network. In DSG-M, the
two physics simulators exchange many balls that are moving at their
border, and that creates a much higher network traffic than in
DSG. Figure~\ref{fig:queuesize} shows messages from the dispatcher
component to one of the physics simulator being increasingly queued as
the physics simulation unfolds. These messages correspond to exchange
of balls (from the other physics simulator) as well as creation of
balls (from the script simulator). The growing queue size meant that
new balls added to the Galton box took longer to arrive, resulting in
less balls to simulate. But as the network got progressively worse, so
did the mean time between creation and deletion. Inadvertently, the
overhead networking cost was masking the scalability results, leading
us to believe that DSG-M was much better than what it was in reality.

These situations, unfortunately, are not uncommon when dealing with
DVEs. The activation of certain behaviors may result in a complex
cascade of effects, some of which may mask others, leading to
erroneous conclusions.

\noindent\textbf{5. Third-party defects.}

Often, the operating environment of DRTs includes a variety of
third-party components, and these can be problematic. In our case,
when we ran experiments under very high load, we ran into a problem
that made the simulators crash at the end of the experiment. We
assumed there was a bug in OpenSimulator or in DSG / DSG-M, and spent
many hours trying to find it. Eventually, we concluded the bug was in
the memory management of the Mono framework~\cite{mono}. By
recompiling a more recent version of Mono with an extra flag, and
using the right garbage collector, the simulators stopped crashing.

These situations, unfortunately, are also not uncommon. Most software,
these days, stands on the shoulders of a very large number of
$3^{rd}$-party frameworks and libraries, over which there is very little
control. When a failure happens, there is a large number of potential
culprits, not just in one's own code, but in the code of the entire
operating environment.

\section{Case Study 2: Login Procedure} \label{sec:testing}

This case study concerns a systematic profiling to assess and control
the impact of user login on OpenSimulator server
performance. OpenSimulator users and developers had observed that user
login was a heavy activity that suffered from lag, but the cause of
lag was unknown. Like in the previous Case Study, we describe the
rationale behind the experiments and metrics, and conclude the section
with observations regarding the challenges encountered.

\subsection{Objective} 

The main objective of this study was to isolate the cause(s) of
perceived lagging (\textit{time sensitivity}), and to mitigate, or
even improve, OpenSimulator once those causes were identified. In this
case, the users' perception of lag correlated very strongly with the
operating system's CPU metric, as high CPU load was always observed
upon user's login. This, in turn, undermined the ability for
OpenSimulator to \textit{scale} to a large number of users during
simultaneous logins, as the server became too busy to be able to
process the login requests within an acceptable time frame.

The secondary objective was to develop test scenarios related to users
login that could be ran automatically, and ensure that future changes
to OpenSimulator would preserve an acceptable behavior upon users login
on the part of the simulator.

\begin{figure}[t]
\center
\includegraphics[width=\linewidth]{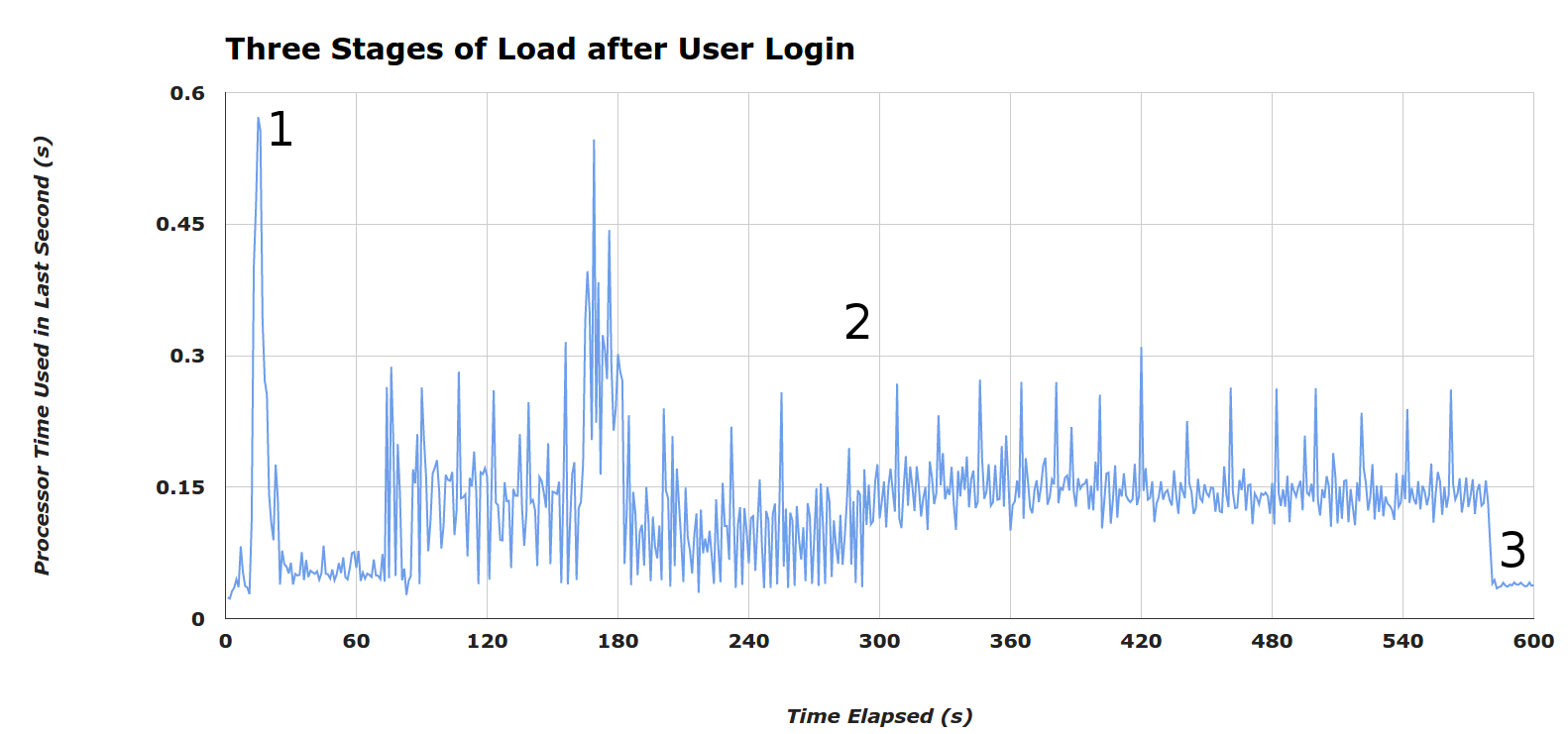}
\caption{Three stages of server performance load during a high-load user login. \textbf{1}: CPU load spike after login; \textbf{2}: Load during inventory and scene; \textbf{3}: return to steady-state levels after login complete.}
\label{fig:loadspike}
\end{figure}

To login to an OpenSimulator simulation server, a user enters a server
URI and credentials into a client viewer. Once the login is
authenticated, the user's client needs to download a large volume of
virtual entities so that its local virtual environment state is
consistent with the server's state. This means that the user entering
the system demands a high volume of resources from the simulation
server, thus limiting bandwidth resources available to other logged in
clients. The client caches many of the assets between sessions, and
this somewhat lowers the login load upon subsequent logins. We were
interested in the login procedure in the worst case scenario, i.e. the
first time the user logs in to a certain virtual space that they have
never visited before. Figure \ref{fig:loadspike} shows privileged
processor time (roughly, CPU\%) accumulated in one example login into
OpenSimulator. Three stages of login are identified in the figure: (1)
an initial load spike, (2) content retrieval, and (3) return to
steady-state levels. The underlying reasons for the initial
performance spike and content retrieval load were unknown.

\subsection{Experiment} 

\begin{table*}
\caption{Login Experiment Configurations}
\label{tab:performancefactors}
\centering
\ra{1.2}
\begin{tabular}{@{}lclclclcl@{}}\toprule
\textbf{Factor} && \textbf{Configuration} && \textbf{Size $^{1}$} && \textbf{File Contents $^{2}$} && \textbf{Graphics}\\
\midrule
Avatar Weight && Light Avatar && 0.33 MB && 136 items && Figure \ref{fig:avatar-a}\\
              && Heavy Avatar && 1.1 MB && 183 items && Figure \ref{fig:avatar-b}\\ 
\hline
Inventory Size && Light Inventory && 0 MB && 0 additional folders and items &&\\
               && Heavy Inventory && 20.6 MB && 8,977 folders with 31,986 items &&\\
\hline
Scene Complexity && Light Scene && 0.038 MB && 2 scene objects, 2 assets && Figure \ref{fig:scenes-a} \\
                 && Heavy Scene && 185.4 MB && 238 scene objects, 1171 assets && Figure \ref{fig:scenes-b} \\
\hline
\multicolumn{3}{l}{\textit{1. All inventory and scene formats gzip compressed}} \\
\multicolumn{3}{l}{\textit{2. Scene contents computed with oarinfo.py utility \cite{clarkcasey2014oar}}}\\
\end{tabular}
\end{table*}

\begin{figure*}
\centering
\subfloat[][]{%
\label{fig:avatar-a}%
\includegraphics[width=0.16\linewidth]{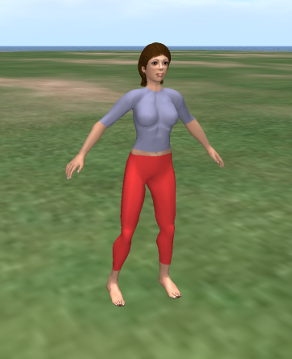}}%
\subfloat[][]{%
\label{fig:avatar-b}%
\includegraphics[width=0.16\linewidth]{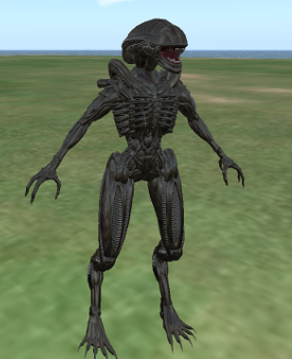}}%
\subfloat[][]{%
\label{fig:scenes-a}%
\hspace{0.2cm}
\includegraphics[width=0.33\linewidth]{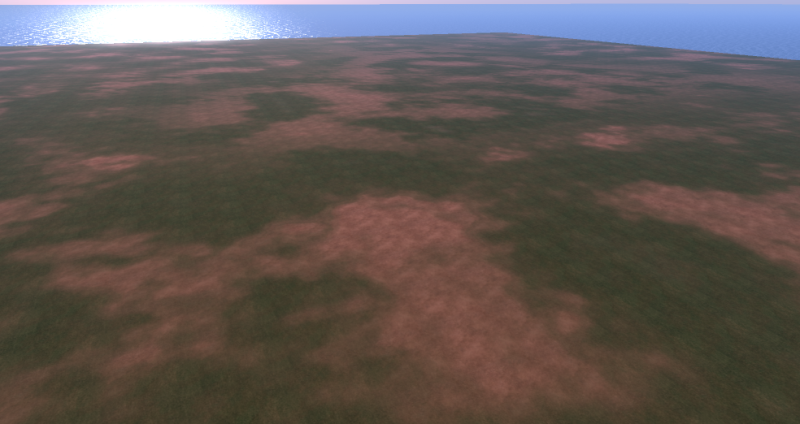}}%
\subfloat[][]{%
\label{fig:scenes-b}%
\includegraphics[width=0.33\linewidth]{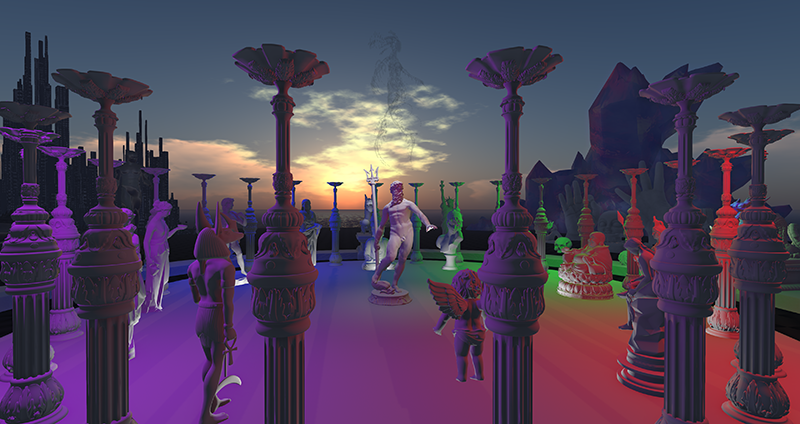}}\\
\caption[]{Avatar and scene configurations from login study. \textbf{\subref{fig:avatar-a}} Ruth, light-weight baseline avatar (left); \textbf{\subref{fig:avatar-b}} Alien, heavy-weight avatar; \textbf{\subref{fig:scenes-a}} Light scene; \textbf{\subref{fig:scenes-b}} Heavy scene \cite{reticulation2014}.}
\end{figure*}

Our in-depth knowledge of OpenSimulator indicated that three factors
might impact server performance during login, namely:

\begin{enumerate}
  \item \textbf{Avatar Weight}: The complexity of a user's appearance,
    including textures, skins, and scripts.
  \item \textbf{Inventory Size}: Inventories are file-folder
    structures containing virtual objects that belong to the
    user. 
  \item \textbf{3D Scene Complexity}: The objects, scripts, textures,
    and meshes contained in an OpenSimulator region.
\end{enumerate}

As such, we designed an experiment meant to measure the effect of each
of these factors, independently, towards CPU
load. Table~\ref{tab:performancefactors} summarizes the experimental
configurations, with a ``light'' and a ``heavy'' configuration for
each factor. Figures~\ref{fig:avatar-a} and \ref{fig:avatar-b} show
the light and heavy avatars; figures~\ref{fig:scenes-a} and
\ref{fig:scenes-b} show the light and heavy scenes. The experiment
consisted in measuring CPU load upon one user's login under the 8
scenarios resulting from the complete combination of configurations,
i.e. light avatar + light inventory + light scene, light avatar +
light inventory + heavy scene, etc.

Possible compounding factors were eliminated. First, the client's
cache was always cleared in between experiments. Second, given
previous work concerning performance impact of avatars in
OpenSimulator~\cite{gabrielova2014impact}, we know that the avatar's
actions in the world have a measurable impact on server load, as
measured by CPU; for example, standing vs. seating have different
performance profiles, because a seated avatar is removed from the
physics simulation; walking vs. standing also have different
performance profiles; etc. In this study, the avatars remain standing
and do not interact with any object on the scene, so that confounding
processing load is minimized.
 
\subsection{Setup} 

OpenSimulator was configured in a ``grid'' configuration, where the
space simulation server is separated from the central resource server
that serves the login request as well as many resources stored in a
database. As such, the set up for the experiments consisted of three
networked components: (1) the client; (2) the space simulation server;
and (3) the central resource server. The architecture of OpenSimulator
is such that the space simulation server always proxies the access to
backend resources, such as inventory and textures -- i.e. the client
never accesses the central server directly, except for the initial
login request.

A single avatar was logged into the region server using the
Singularity Viewer~\cite{singularity2014}, an open source client for
OpenSimulator. The viewer was configured with high graphics quality
and a large draw distance to simulate a full view of the virtual
scene. Each experiment run lasted for 600 seconds (10 minutes),
starting from the time at which the region received a login request
for the user. This time was chosen because most inventory
configurations loaded within 10 minutes. Five experiment runs were
performed for each combination of avatar weight, inventory size, and
scene complexity --- 40 runs in total. The built-in OpenSimulator
logging and monitoring utilities were used to record data.

The experiments were conducted on wired machines on the UC Irvine
network. Test avatars were logged in to the OpenSimulator server from
a laptop with Intel i5-2520M CPU (2.50GHz), 4 GB of RAM, and an Intel
integrated graphics card. The OpenSimulator simulation servers in this
study were hosted on a Dell machine with an Intel i5-4670 CPU (3.40
GHz) with 4 cores and 8 GB of RAM. The machine ran the Ubuntu 12.04
LTS operating system. Monitoring and statistics logging occurred on
this machine.

\subsection{Metrics and Results} 

\begin{figure*}%
\centering
\includegraphics[width=0.85\linewidth]{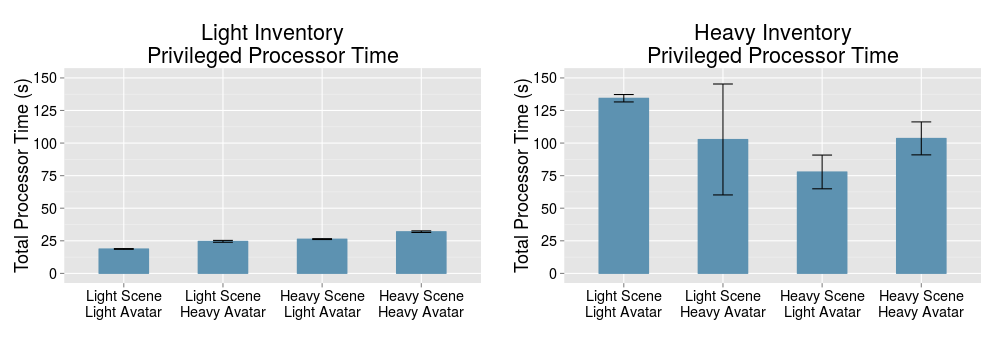}
\caption{Simulation server privileged processor time, comparing light inventory (left) and heavy inventory (right) configurations.}
\label{fig:login_compare_inventories}
\end{figure*}

The primary metric was accumulated privileged processor time used by
the simulation [space] server. This metric is computed within
OpenSimulator's monitoring component, and it measures CPU usage by the
server process over time. Two metrics were later collected for debugging
purposes: the quantity of inventory folder requests received by the
server, as well as the quantity of HTTP packet requests received by
the server.

After conducting five tests for each combination of login
configurations (6.5 hours of combined tests), it appeared that the
size of the user inventory had the highest impact on performance
load. Figure~\ref{fig:login_compare_inventories} shows average
load measurements between configurations with light inventories and
heavy inventories. We observed that configurations with heavy
inventories resulted in many server requests for nested inventory
folders. The impact of the avatar complexity seemed to be
negligible. The scene complexity had some effect, but not as much as
inventory size. We also observed a puzzling performance profiles in
two of the experimental configurations that are discussed next.

In order to further study the performance issue with inventory, we
made additional experiments where we added a fourth component to the
setup, specifically a dedicated inventory server. This server was
configured to handle all inventory folder requests directly from the
client, meaning that inventory retrieval was no longer proxied by the
simulator server. The goal of this additional experiment was to verify
whether removing inventory service altogether from the simulation
server would bring CPU usage to acceptable levels at the simulator, or
whether there was something more complex going on. We conducted all
experimental runs again with the added component in the experimental
setup.

\begin{figure*}%
\centering
\includegraphics[width=0.85\linewidth]{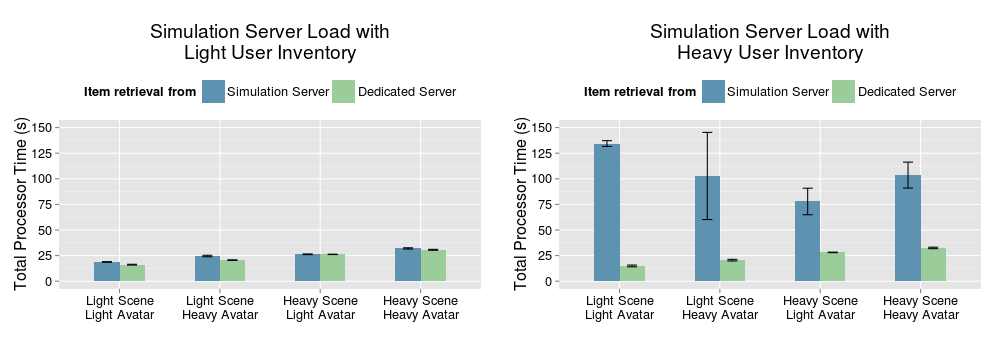}
\caption{Simulation server privileged processor time with a dedicated simulation server for inventory, comparing light inventory (left) and heavy inventory (right) configurations.}
\label{fig:login_compare_server_setups}
\end{figure*}

Indeed, adding a dedicated server for inventory retrieval reduced
server load to levels comparable to those obtained for light
inventories. Figure~\ref{fig:login_compare_server_setups} shows this
additional result. This meant that inventory servicing at the
simulation server was the sole cause of the observed high CPU
usage. That, in turn, gave us the necessary confidence to start
solving the problem by looking at the code that handled inventory
servicing at the simulator. We found very problematic code, changed
it, and eventually fixed all these issues with initial inventory
downloads.

\subsection{Observations}

\noindent\textbf{1. Time sensitivity.}

The experiment was designed to profile a specific user activity
(login) that consistently showed lag for users. ``Lag'' is an informal
term used in DVEs that denotes situations where the interaction with
the environment \textit{feels} slower than expected. As such, by
definition, lag is a perceptual phenomenon; it may correlate with
system-level metrics in complex ways, or not at all.

In this case, there was a very strong correlation between the lag felt
by users and CPU usage at the simulation server. Clearly, the code
executed at login was making the CPU busy. By focusing on measuring
CPU and, eventually, decrease its usage during login, we hoped to
decrease the lag felt by users. This was a hope that might or might
not come to fruition, as lag and CPU usage are not the same thing.

After these profiling experiments, we improved the login code of
OpenSimulator considerably, reducing CPU usage to a small fraction of
what was measured in these experiments. Fortunately, these
improvements resulted in a considerable reduction of lag too. We
measured this reduction in lag qualitatively, by releasing the
fixes made to OpenSimulator to the community and requesting feedback
from them.\footnote{See http://opensimulator.org/mantis/view.php?id=7564}

Similarly to observations made regarding correctness of simulation in
Case Study 1, the user experience is the most important aspect DVEs,
and that is, ultimately, what needs to be measured. However, not only
user experience is hard to measure, but it becomes impractical to
measure it \textit{while} developing these systems. For example, it
would have been highly ineffective to ask independent subjects to
check the lag after every important code commit that seemed to reduce
the CPU usage. System-level metrics are much easier to measure, but
they might or might not correlate with the observable effects that
matter to users.

\noindent\textbf{2. Non-functional defects.} 

This study exposed two puzzling performance profiles, all related to
heavy inventory configurations -- see
Figure~\ref{fig:login_compare_inventories}, right bar chart. One of
them pertained to the configuration heavy inventory + heavy avatar +
light scene. Those experiments had a very wide variation in CPU usage,
as seen in the error line (second bar from left). The second one
pertained to the configurations heavy inventory + light avatar: the
experiments with the light scene (first bar from the left) showed
higher CPU usage than the experiments with the heavy scene (third bar
from the left); this was counterintuitive.

After measuring a couple of other internal quantities, we concluded
that both of these situations could only be explained by the existence
of bugs in the code of OpenSimulator. However, these weren't
functional defects related to \textit{correctness} of behavior, as the
function performed by the simulator was essentially correct --
inventory was downloaded by the client, eventually. These were defects
affecting the non-functional properties of OpenSimulator. In one case,
the CPU load was very unpredictable; in the other, something was
making the CPU unreasonably busy on light scenes when compared to
heavy scenes.

Non-functional defects are much harder to deal with than functional
defects. First, a specification usually does not exist upfront, not
even an upfront expectation of correct behavior; ``I know something is
wrong when I see it,'' seems to be the main approach to identifying
these issues. For example, we can define upfront the inventory
download feature, as that is a fundamental part of the login
procedure, but it is much harder to define upfront the non-functional
property related to variance in CPU usage of inventory download,
because it is one of a possibly unbounded list of non-functional
behaviors. Second, non-functional defects may show up only when
certain conditions are met, making them very hard to reproduce. For
example, while we were fixing the inventory download issue, it became
apparent that the distance between the simulation server and the
central server had a significant impact on this defect, something that
caused a fair amount of confusion; for a certain OpenSimulator grid
whose central server is hosted in a US data center, some users in
Europe experienced issues that users in the US could not
reproduce.\footnote{See, for example,
  http://opensimulator.org/mantis/view.php?id=7567} Finally, once the
non-functional defects become apparent, it is much harder to develop
regression tests for non-functional properties, such as those exposed
by these two puzzling performance profiles.

While all software is affected by non-functional properties, DVEs, and
DRTs in general, are particularly exposed to them, as the existence of
distributed components and the expectation of real-time interaction
pose difficult challenges in terms of identifying non-functional
defects, reproducing them, and making sure they do not come back once
they are fixed.

\noindent\textbf{3. Masking, again.} 

Distributed non-deterministic interaction between components may lead
to masking of known and unknown incorrect behaviors. We encountered this
already in Case Study 1, when queues acted as buffers of the balls and
spared CPU from having the expected load on the physics
simulators. Here, too, we observed masking of incorrect
[non-functional] behavior when we moved the inventory service to a
separate server. By doing that, we eliminated the abnormalities
described above, but the defects were still there in the code. They
just became invisible.

As we fixed the code, it was clear that the API of the inventory
service was highly inefficient. When the calls were on the same
component (such as in the case of a dedicated server) those
inefficiencies were not noticeable; however, when the calls came
from a component on the network, those inefficiencies became visible,
and produced the results we measured in these experiments. 

\noindent\textbf{4. Automation.} 

The profiling of any software requires its execution, as well as the
triggering of specific inputs. Our profiling experiments were
labour-intensive: for each of the 45 measurements we had to start 3
components (4, in the case of the dedicated inventory server) and then
login a user manually. This was clearly not ideal, but we had no other
option: we didn't know where the cause of the high CPU was, and it
could partly be the graphical client, Singularity -- a very large and
complex piece of software that we treat as a black box and for which
there is no headless version.

Once we were confident that the cause was a non-functional bug in
OpenSimulator, we then replaced the graphical client with a very
simple headless client that only downloaded the inventory. This
allowed us to reproduce the high CPU usage at the simulator without
having to deal with the graphical client. But the process of starting
and stopping components was still manual. Also, we have not been able
to add any regression tests regarding this issue, as that would
require a framework for distributed testing of non-functional defects
that OpenSimulator does not have.

It would be desirable to develop a framework for automatic testing of
these specific distributed components, particularly tuned for testing
application-specific non-functional behaviors, but that is challenging
goal. To the best of our knowledge, that does not exist.

\section{Historical Perspective}
\label{sec:related}

This section examines the challenges of DVEs in greater depth, and
offers a broader historical context for the observations we made about
our experiments. Many of the challenges we encountered in these two
case studies have been analyzed in the literature. The goal of this
historical perspective section is to argue that the difficulties in
designing, testing and evaluating DVEs are inherent, and not just the
consequence of inept engineering. New ideas for addressing them are
needed.

\subsection{On Design and Evaluation}
\label{subsec:related-design}

We discuss the unique characteristics of DVEs that impact their design
and evaluation.

\noindent\textbf{On Correctness}

Since the early days of computing, \textbf{correctness} has had a
well-established definition: an algorithm, or a system, is correct if
it honors its specification (see,
e.g.~\cite{gupta1992,prasad2005}). \textit{Functional} correctness
pertains to the input-output behavior of the algorithm or the
system. This commentary focuses on functional correctness, but we use
the word ``correctness'' for brevity.

The size and complexity of what can be proven correct has been growing
at a steady pace~\cite{prasad2005}, and it is conceivable that in the
future extremely complex systems like OpenSimulator could be formally
specified and verified; we are still a long way from that.  It is not
our intention to cover the impressive progress of formal verification
techniques of recent years, including for real-time and distributed
systems~\cite{kwiatkowska2011,loos2011,garavel2013}. Instead, we want
to discuss the definition of correctness provided above, how it
interferes with other design concerns of DVEs, and how researchers and
developers have been coping with those interferences.

In discussing Case Study 1, we mentioned that correctness of a DVE can
be seen under two perspectives: the user and the function itself. In
the case of a physics simulation, like in our experiments, it is
desirable that it is as realistic as possible -- ideally simulating
exactly physics in the real world. But such goal carries with it a
heavy demand on computing power, and it becomes unachievable in
practice. In order to keep the simulation's performance under control,
developers of physics simulation engines simply make better-performing
numerical approximations of real physics everywhere they can
(see e.g.~\cite{battaglia2013}). In doing so, the simulation deviates
from the correct behavior. As observed in Case Study 1, when
simulating a Galton Box, the physics engine in OpenSimulator produces
a result that does not match the correct functional behavior dictated
by the laws of physics in the real world, which led us to having to
use an empirical baseline.

Physics is not the only aspect of DVEs that is subject to
correctness-degrading approximations. Performance and responsiveness
of these systems usually take precedence over correctness. This gives
rise to a different, and more recent set of metrics for assessing
correctness, those that are based on human
perception~\cite{Reitsma:2003,Yeh:2009,Lin:2011,Corsini:2013}. The
basic premise of this line of work is this: if humans cannot tell the
difference between an exactly correct behavior and an incorrect
behavior that requires less computing resources, then the latter is
preferred. This makes up for substantially different DRT systems than
those traditionally envisioned in DRT research.

\noindent\textbf{On Time Sensitivity and Consistency}

In Case Study 2, we mentioned that users of OpenSimulator reported lag
upon certain logins. In OpenSimulator, lag is usually felt in terms of
physics -- e.g. walking (of self and others) is not smooth, collision
detection is slower than expected -- and in terms of the environment's
response to their inputs -- e.g. clicking on an object to trigger some
effect produces that effect much later than expected. The workload generated by graphics rendering and event propagation can lead to latencies over 3 times larger than video streaming \cite{roberts2009}.

Real-time systems need to perform operations in a time sensitive
manner. Sha et al.~\cite{sha2004real} give a comprehensive overview of
the history of the major developments in real-time scheduling,
including for distributed systems; we refer the reader to that
interesting paper for the historical perspective on dealing with time
sensitivity in computing systems. DVEs, in particular, are designed
under the expectation that their responsiveness matches, to some
approximation, the speed of interactions in the real
world~\cite{stytz1996distributed}. However, unlike hardware control
DRT systems such as anti-lock brakes or the control of a rocket, DVEs
are interactive systems ultimately used by people; as such, and as
mentioned before for correctness, it is important to include the human
perceptual system as a parameter of the design and assessment. This
inclusion has two consequences for design: compensation for delays and
variance, and opportunity for optimizations.

On the one hand, the network introduces latency and jitter. This has
been known for a long time in the engineering of DVEs; well-known
solutions to these problems include efficient server placement algorithms \cite{duong2008} and several prediction techniques such
as the centuries' old dead reckoning~\cite{solver1958dead} applied to
these environments~\cite{pantel2002suitability, hakiri2010}. These techniques help
in improving responsiveness and in preserving the illusion of
consistency among the distributed components, even if the system is
not exactly consistent.\footnote{See
  e.g.~\cite{armitage2006networking} for a good overview of latency
  compensation techniques.}

On the other hand, the perceptual effects of latency and jitter have
been studied more recently in the research literature with the goal
towards devising optimizations that improve the perceived
responsiveness of
DVEs~\cite{Quax:2004,Dick2005,gupta2008semmo,chen2011perceptual}.

Just like for correctness, this line of work in perceptual metrics is
very important for DVEs, as many more opportunities for optimization
will likely be found that will make these systems more
scalable. However, equally important is the mapping of such metrics to
system-level metrics, as user studies carry an unbearable overhead
during development of the system.

\noindent\textbf{On Scaling and Overhead Costs}

Jim Waldo stated~\cite{waldo2008scaling}: ``Online games and virtual
worlds have familiar scaling requirements, but don't be fooled:
everything you know is wrong.'' This is an overstatement, but it is
true that the focus on user experience in DVEs requires us to rethink
many of the concepts we took for granted regarding DRTs. 

A large virtual environment is usually associated with a large
workload. If the virtual environment has thousands of users and
objects, computing the result of each interaction at a every time step
is unfeasible within the limited time frame required for reasonable
interactivity, usually in the hundreds of milliseconds. DVEs typically
partition this workload into multiple simulators by virtual space,
with simulators being responsible for unique areas of the virtual
environment. This idea can be traced back to
Locales~\cite{barrus1996locales} and DIVE~\cite{frecon1998dive}, and
it has been the main architectural approach to scalability of massive
multi-user environments.

Many improvements and variations of this idea have been proposed over
the years. For example, microcells of custom size and shape add
flexibility to adapt the load among machines
dynamically~\cite{Vleeschauwer}. A push-pull framework can be used to filter and reduce the number of messages exchanged between partitions \cite{minson2007}. Another variation, sharding, is a
form of space partitioning where different users connect to different
copies of parts of the space.\footnote{Unfortunately, the origin of
  the word \textit{sharding}, and corresponding technique, seems to
  have been lost in history. It likely came from the game Ultima
  Online launched in 1997, which may have been the first one to
  provide multiple copies of game spaces for different groups of
  users.} 

Load partitioning among servers is, therefore, the only way a virtual
environment can scale. However, space is not the only aspect of these
environments that can be partitioned. The DSG architecture, for
example, partitions the workload by functionality, such as physics and
script execution~\cite{Lake2010,Liu2010,Liu:2010}. Project
Darkstar~\cite{waldo2008scaling} divides the load by task triggered by
user input. Kiwano~\cite{diaconu2013kiwano} divides the world by
space, and groups the load generated by the users' input by their
proximity in space. Meru~\cite{horn2009scaling} partitions the load by
both space and content (3D objects). 

The art of designing scalable DVEs lies in finding the right load
partitions for the purpose for which the DVE is being designed. Load
partitions carry overhead costs in terms of coordination among
servers. In a ``good'' partition design, the system will scale
horizontally, i.e. more load can be handled by adding more servers in
as linear a relation as possible; in a ``bad'' partition design the
benefits of load distribution will be dwarfed by the communication
overhead among servers. For example, in our worst-case scenario
experiments in Case Study 1, only a 15\% improvement was observed when
dividing the space in two; nearly 85\% of the computation was
being used for the overhead of synchronizing the simulators. The
overhead was mostly due to object migration causing numerous messages
related to the creation and deletion of tens of thousands of objects.

Finding the appropriate load partitions for a DVE requires a
considerable amount of experimentation, of the kinds we did for
DSG-M. Ideas that look good on paper often fall short when placed into
an actual system. Scalability in DVEs is still very much a topic of
research. One of the confounding factors in this research area is the
absence of common objectives among the different systems. They all
claim to scale, but the applications for which they are being
designed, and therefore the metrics they use to assess scalability,
are all very different. 

In general, the volume of concurrent user interactions and the
complexity of the shared artifacts are the two most important factors
governing approaches to scalability
~\cite{kumar2008second,kinicki2008traffic,Liu:2010,dionisio2013virtual,singh2014controlling}. Given
that different applications have very different demands of user-user
and user-environment interactions, it follows that common metrics are
hard to find. The development of DVEs has been primarily driven by
commercial interests and realized by skillful engineers. Research in
these systems, so far, has been fairly ad-hoc. Calls for making it
more systematic are now starting to appear~\cite{singh2014controlling}.

Giving a positive spin to Waldo's observation, there is a fair amount
of work to be done in categorizing dimensions of scalability in DVEs,
and in producing benchmarks and metrics that can be used to compare
different solutions for the same categories of problems.

\noindent\textbf{On Faults, Fault-Tolerance and Fault Prevention}

The concept of \textbf{fault tolerance} emerged a long time ago, and
for a while, it remained associated with hardware
design~\cite{pierce1965}. As software became more pervasive and
important, those same ideas were adopted for software
systems~\cite{randell1975}. For a long time, software fault-tolerance
meant almost exclusively the existence and management of ``stand-by
sparing'' components; it eventually grew to encompass a much larger
scope of concepts and techniques, such as fault detection and fault
models (see,
e.g.~\cite{cristian1991understanding,isermann2006}). Distributed
systems, in particular, are fundamentally
unreliable~\cite{muir2004seven}. Our case studies do not illustrate
the traditional concepts of fault tolerance very well, those where
possible faults are known in advance and mitigated in some way; but
the effect of non-functional, unanticipated software faults was
prevalent in our observations.

The existence of these unanticipated faults has been discussed in the
literature for a long time. For example, as early as 1990 Lee and
Anderson wrote in their textbook~\cite{lee1990}:

\begin{quotation}
``While the anticipation of faults has been successful in the past for
hardware, the present trend towards very large scale integration is
already casting doubt on the validity of the assumption that all
component failure modes are known.[...] It follows that unanticipated
faults are likely to arise in hardware systems, and will certainly
require tolerance in high reliability applications.

However, there is a much more important and insidious reason for the
occurrence of unanticipated situations. If there are faults in the
design of a system then the effects of such faults will be
unanticipated (and unanticipatable!) [...]

While design faults may have been uncommon in hardware systems, the
only type of fault that can be introduced into a non-physical system,
such as a system implemented in software, is a design fault
[...]. Applications of the fault prevention techniques that have been
successful for exposing design faults in hardware systems have only
met with limited success when applied to software.''
\end{quotation}

This text is as valid today as it was in 1990. Since then, a great
deal of progress has been made in capturing functional behavior of
software. The area of software testing, for example, has seen an
enormous growth (more on this later on). More recently, the area of
formal specification and verification has also gained considerable
attention. 

Unfortunately, very little progress has been made in formally
identifying and capturing non-functional behaviors, such as those we
described, to the point of preventing non-functional defects from
occurring. Additionally, very little progress has been made in formally
identifying and capturing operational expectations from $3^{rd}$-party
components to the point of preventing defects in these components from
causing our own systems to fail. DVEs, and DRTs in general, are
particularly exposed to these two kinds of hard software engineering
problems, as they tend to rely on a large stack of $3^{rd}$-party
components, and a considerable portion of their existence is
determined by \textit{how} they do the things they do (i.e. operation
rather than function).

A lot more needs to be done in addressing the unavoidable existence of
{\bf design} faults, especially in what concerns non-functional
aspects of the designs.

\subsection{On Testing}

When discussing the non-functional defect studied in Case Study 2, we
mentioned that one unfulfilled goal of that work was to write some
sort of regression test that would ensure that specific unreliable
behavior with inventory download will not come back again. Regression
tests are standard practice in software development, and
OpenSimulator has hundreds of them. Unfortunately, when it comes to
identifying and testing the occurrence of known non-functional
defects, especially in distributed systems, the research literature is
scarce.

The research field of software testing can be traced back to the early
1970s (for a good overview, we refer the reader
to~\cite{bertolino2007software}). Most testing research and
development has focused on functional, non-distributed testing. Once
described as a ``lost world of debugging and
testing''~\cite{glass1980real}, distributed systems always lagged
behind relevant developments in testing and debugging. Unfortunately,
the situation has not improved much: while the area of testing has
seen enormous progress in the last 30 years~\cite{orso2014software},
distributed systems testing is still lagging behind non-distributed
systems testing. Unit testing frameworks, code coverage
frameworks, fault injection and all sorts of test input generators
have made the transition from research to practice, but up to today,
we continue to see research literature refer to testing of distributed
systems as ``a challenge''~\cite{gupta2011diecast,canini2011toward}.

The challenges in distributed real-time systems testing were first
clearly formulated by Sch\"utz in a paper that is still as relevant
today as it was in 1994~\cite{schutz1994fundamental}. This work
identified six fundamental issues in DRT testing: organization of test
phases; observability of DRT systems; reproducibility to cope with
non-deterministic behavior; splitting testing between development
hosts and target systems, environment simulation to support real-time
correctness, and representativity of realistic inputs. Sch\"utz's
paper is still a solid blueprint not just for the challenges of
testing DRT, but it also inspires potential solutions for those
challenges.

Indeed, in recent years there has been some progress in distributed
systems
testing~\cite{rutherford2008evaluating,dao2009,canini2011toward}. The
most recent advances in software testing for distributed systems
include model checking (explored in depth
in~\cite{killian2007life,lin2009modist}) and capture-replay
testing~\cite{joshi2007scarpe}, which can be traced back
to Tsai's work in non-intrusive real-time system testing and debugging
\cite{tsai1990noninterference}.

We note, however, that our faulty inventory download behavior does not
fit well in Sch\"utz's framework, specifically in what concerns
reproducibility. We know that part of the inventory bug was
non-deterministic; its non-determinism was not about the function
itself (there was no functional defect) but about CPU
usage. Furthermore, even though the defect was non-deterministic, it
is possible to describe it using a statistical specification: part of
the bug was that it produced a wide variation in CPU usage between
independent login sessions. Clearly, if that happens again after our
fixes, it means that the there is a regression. Sch\"utz's description
of the reproducibility concern did not take into account the existence
of these kinds of tests that require launching the execution several
times and observing the statistics of some metric. We believe that
repeated experimentation is an important aspect of testing DVEs and
DRTs in general, especially when it comes to testing non-functional
properties.

This leads us to the most relevant literature for the testing problems
we describe in our case studies: testing non-functional properties of
software. A paper by Weyuker et al. published in 2000 notices the
almost complete lack of literature related to performance testing at
the time~\cite{weyuker2000experience}, contrasted with how important
the problem was/is in industrial projects. They then describe a case
study within AT\&T, explaining how performance tests were
specified. This included having to collect data in order to establish
realistic workloads for the tests -- similarly to our profiling work
and the setup of the tests. They also make this insightful
observation:

\begin{quotation}
``It is essential to recognize that designing test suites for
performance testing is fundamentally different from designing
test suites for functionality testing. When doing
functionality testing, the actual values of parameter settings
is of utmost importance. For performance testing, in
contrast, the primary concerns are the workloads and
frequencies of inputs, with the particular values of inputs
frequently being of little importance.''
\end{quotation}

Since then, there have been a few more papers on the topic. 
Denaro et al.~\cite{denaro2004early} propose deriving and running
performance tests from early architectural decisions, even before the
software is written. Their assumption is that ``[...]  performance
measurements in the presence of the stubs are good enough
approximations of the actual performance of the final application.''
They provide some preliminary evidence regarding one use case of one
architecture in a J2EE tutorial, but this assumption is difficult to
accept in general.

In a study of unit testing non-functional properties of distributed
systems~\cite{hill2009unit}, Hill et al. observe that many
non-functional DRT properties, not just performance, are often
relegated to system integration stages of software development, and
that conventional system execution modeling tools do not provide the
necessary support for unit testing non-functional requirements. They
then propose their approach to testing non-functional properties,
which is based on mining logs.

As Weyuker et al. noticed in 2000, the lack of research literature on
the subject does not reflect the importance of the subject in
industry, especially for Web applications. That importance can be seen
in the existence of several developer-oriented books on the subject,
e.g.~\cite{erinle2013,molyneaux2014}. These books tend to focus on the
practicalities and tools of performance testing, and do not offer
suggestions for how to advance the state of the art.

\section{Conclusion} 
\label{sec:conclusion} 

Distributed Virtual Environments (DVE) are Distributed Real-Time (DRT)
systems designed with the general goal of connecting multiple users
over the Internet instantly with each other and with a shared virtual
space. DVEs inherit some of the intrinsic difficulties of DRT systems,
such as the overhead of distribution and the overarching importance of
responsiveness and performance. They also have unique characteristics
that make them different from traditional DRT systems. Specifically, a
strong focus on user experience, and the quality of that experience,
requires a re-evaluation of some of the concepts in traditional DRTs.

This paper first presented two case studies describing design and profiling
work previously done by us on one DVE, OpenSimulator. OpenSimulator is
an open-source virtual environment framework that uses the same
protocol as Second Life, and, as such, supports Second Life-like
environments. The first case study focused on assessing whether a
specific  design idea we had was beneficial or not. The
second case study focused on profiling a specific activity (user
login) that OpenSimulator users had reported to suffer from poor
responsiveness (i.e. lag), with the hope of fixing it, and then making
sure it would not regress with future changes to OpenSimulator. In
doing both of these pieces of work, we encountered many challenges
related to metrics and their interpretation, baselines, dependent
variables and masking of defects, non-functional defects, and
automation. We described and discussed these challenges for each case
study.

In order to place our observations in a broader context, and to show
that they represent foundational challenges in DVEs, we then presented
an historical perspective on the design and testing of DVEs and DRTs,
focusing on the pain points encountered during our experimental work.

We believe our experience with the development of OpenSimulator, and
the placement of that experience in a broader context, shed some
light into the open challenges of DVEs, and the kinds of problems that
are worth solving.

\bibliographystyle{IEEEtran}
%
\bibliography{dsrt}
\end{document}